%% file: TOP-11-022_temp.tex
\pdfoutput=1

\documentclass[11pt,twoside,a4paper,cmspaper,final,collab]{cms-tdr}
\def\svnVersion{177021}\def\cmsCernNo{2012-266}\def\cmsCernDate{\today}\def\cmsMessage{Submitted to Physical Review Letters}

\begin{document}\cmsNoteHeader{TOP-11-022}

\hyphenation{had-ron-i-za-tion}
\hyphenation{cal-or-i-me-ter}
\hyphenation{de-vices}

\RCS$Revision: 156658 $
\RCS$HeadURL: svn+ssh://svn.cern.ch/reps/tdr2/papers/TOP-11-022/trunk/TOP-11-022.tex $
\RCS$Id: TOP-11-022.tex 156658 2012-11-06 02:22:38Z alverson $
\ifthenelse{\boolean{cms@external}}{\providecommand{\suppMaterial}{thesupplemental material [URL will be inserted by publisher]}}{\providecommand{\suppMaterial}{Appendix~\ref{app:suppMat}}}
\newcommand{\mtop}{$m_{\cPqt}$}
\newcommand{\DY}{\cPZ/$\Pgg^\star$\xspace}
\ifthenelse{\boolean{cms@external}}{%
\newcommand{\scotchrule[1]}{\centering\begin{ruledtabular}\begin{tabular}{#1}}
\newcommand{\donescotchrule}{\end{tabular}\end{ruledtabular}}
\newcommand{\cmsBreak}{\relax}
}{
\newcommand{\scotchrule[1]}{\centering\begin{tabular}{#1}\hline\hline}
\newcommand{\donescotchrule}{\hline\hline\end{tabular}}
\newcommand{\cmsBreak}{\linebreak[4]}
}%%%%%%%%%%%%%%%  Title page %%%%%%%%%%%%%%%%%%%%%%%%

\cmsNoteHeader{TOP-11-022} % This is over-written in the CMS environment: useful as preprint no. for export versions
\title{Evidence for associated production of a single top quark and W boson in pp collisions at \texorpdfstring{$\sqrt{s}$ = 7\TeV}{sqrt(s) = 7 TeV}}% Force line breaks with \\

\date{\today}

\abstract{
Evidence is presented for the associated production of a single top quark and $\PW$ boson in pp collisions at $\sqrt{s}$ = 7\TeV with the CMS experiment at the LHC. The analyzed data corresponds to an integrated luminosity of 4.9\fbinv. The measurement is performed using events with two leptons and a jet originated from a $\cPqb$ quark. A multivariate analysis based on kinematic properties is utilized to separate the $\ttbar$  background from the signal. The observed signal has a significance of $4.0\,\sigma$ and corresponds to a cross section of $16^{+5}_{-4}$\unit{pb}, in agreement with the standard model expectation of $15.6\pm0.4^{+1.0}_{-1.2}$\unit{pb}.
}

\hypersetup{%
pdfauthor={CMS Collaboration},%
pdftitle={Evidence for associated production of a single top quark and W boson in pp collisions at 7 TeV},%
pdfsubject={CMS},%
pdfkeywords={CMS, physics, TOP}}

\maketitle %maketitle comes after all the front information has been supplied

Electroweak production of single top quarks has been first observed by the D0~\cite{Abazov:1165293} and CDF~\cite{Aaltonen:1165362} experiments at the Tevatron. Single-top-quark production proceeds via three processes: the $t$-channel exchange of a virtual $\PW$ boson, the $s$-channel production and decay of a virtual $\PW$ boson, and the associated production of a top quark and a $\PW$ boson (\cPqt\PW). The latter channel, which has a negligible production cross section at the Tevatron, represents a significant contribution to single-top-quark production at the Large Hadron Collider (LHC). Associated $\cPqt\PW$ production is a very interesting production mechanism because of its interference with top quark pair production~\cite{Frixione:2008yi,Belyaev:1998dn,White:2009yt}, its sensitivity to new physics~\cite{PhysRevD.63.014018, Cao200750, PhysRevD.81.034020} and its role as a background to SUSY and Higgs searches. The ATLAS and Compact Muon Solenoid (CMS) experiments have measured the cross section for $t$-channel production~\cite{Aad:2012ux, Chatrchyan:1359133}  while evidence for $\cPqt\PW$ associated production has been presented by the ATLAS experiment~\cite{Aad:2012dj}. This Letter presents the first study from the CMS experiment of $\cPqt\PW$ production in $\Pp\Pp$ collisions at $\sqrt{s} = 7$\TeV.

	\begin{figure}[h]
	  \begin{center}
	    \includegraphics[width=0.3\columnwidth]{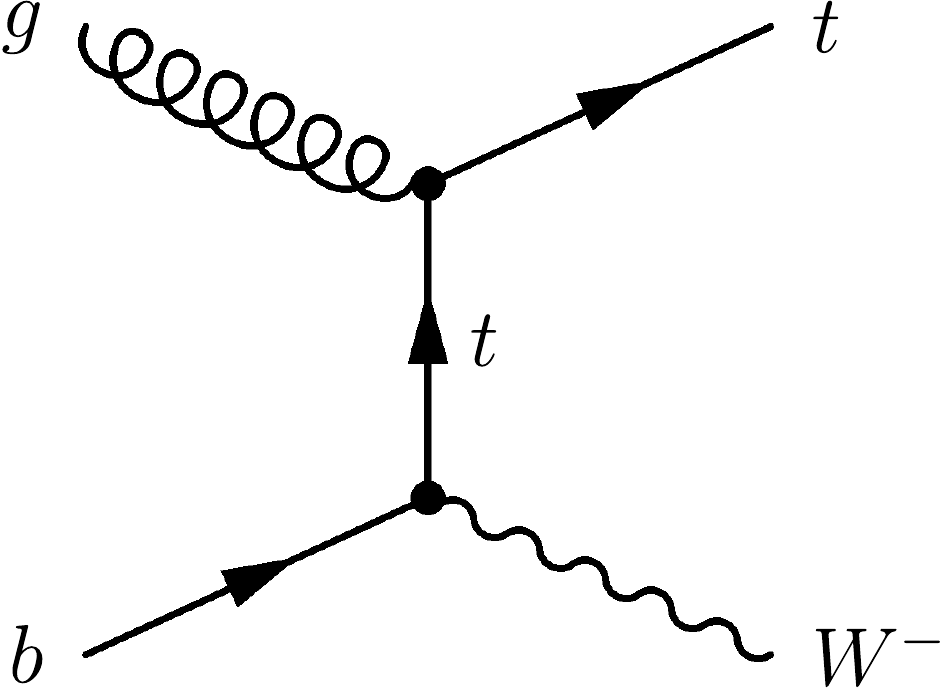}
	    \includegraphics[width=0.3\columnwidth]{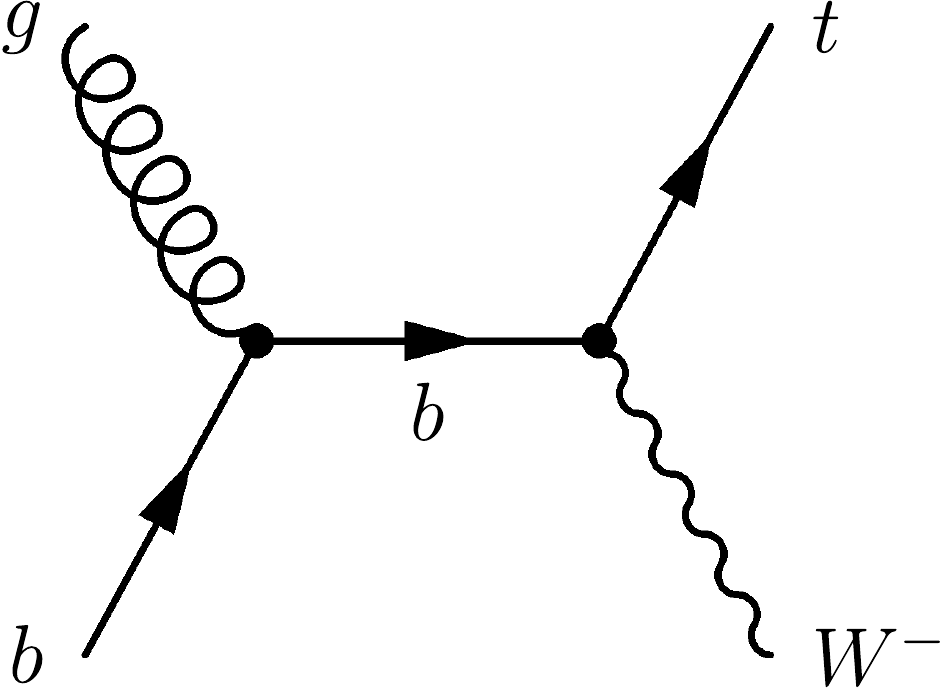}
	    \caption{\label{fig:fdlo}Leading order Feynman diagrams for single-top-quark production in the $\cPqt\PW$ mode, the charge-conjugate modes are implicitly included.}
	  \end{center}
	\end{figure}

The production cross section for $\cPqt\PW$ has been computed at approximate next-to-next-to-leading order (NNLO), the theoretical prediction of the cross section for $\cPqt\PW$ in $\Pp\Pp$ collisions at $\sqrt{s}$ = 7\TeV, assuming a top-quark mass (\mtop) of 172.5\GeV, is $15.6\pm0.4^{+1.0}_{-1.2}$\unit{pb}~\cite{Kidonakis:2010ux}, the first uncertainty corresponds to scale variation and the second to parton distribution function (pdf) sets.

\begin{figure}[h]
	  \begin{center}
	    \includegraphics[width=0.3\columnwidth]{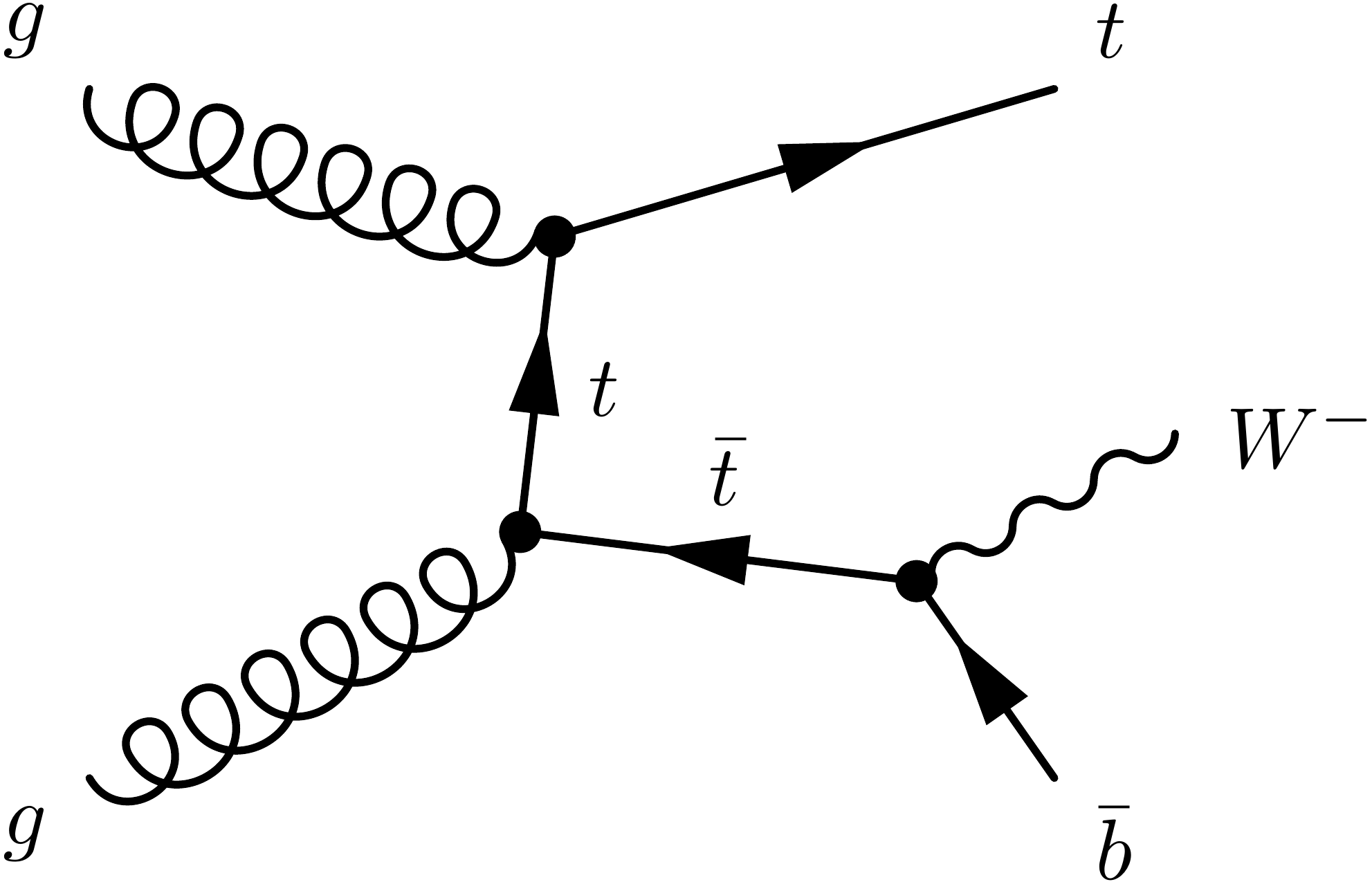}
	    \includegraphics[width=0.3\columnwidth]{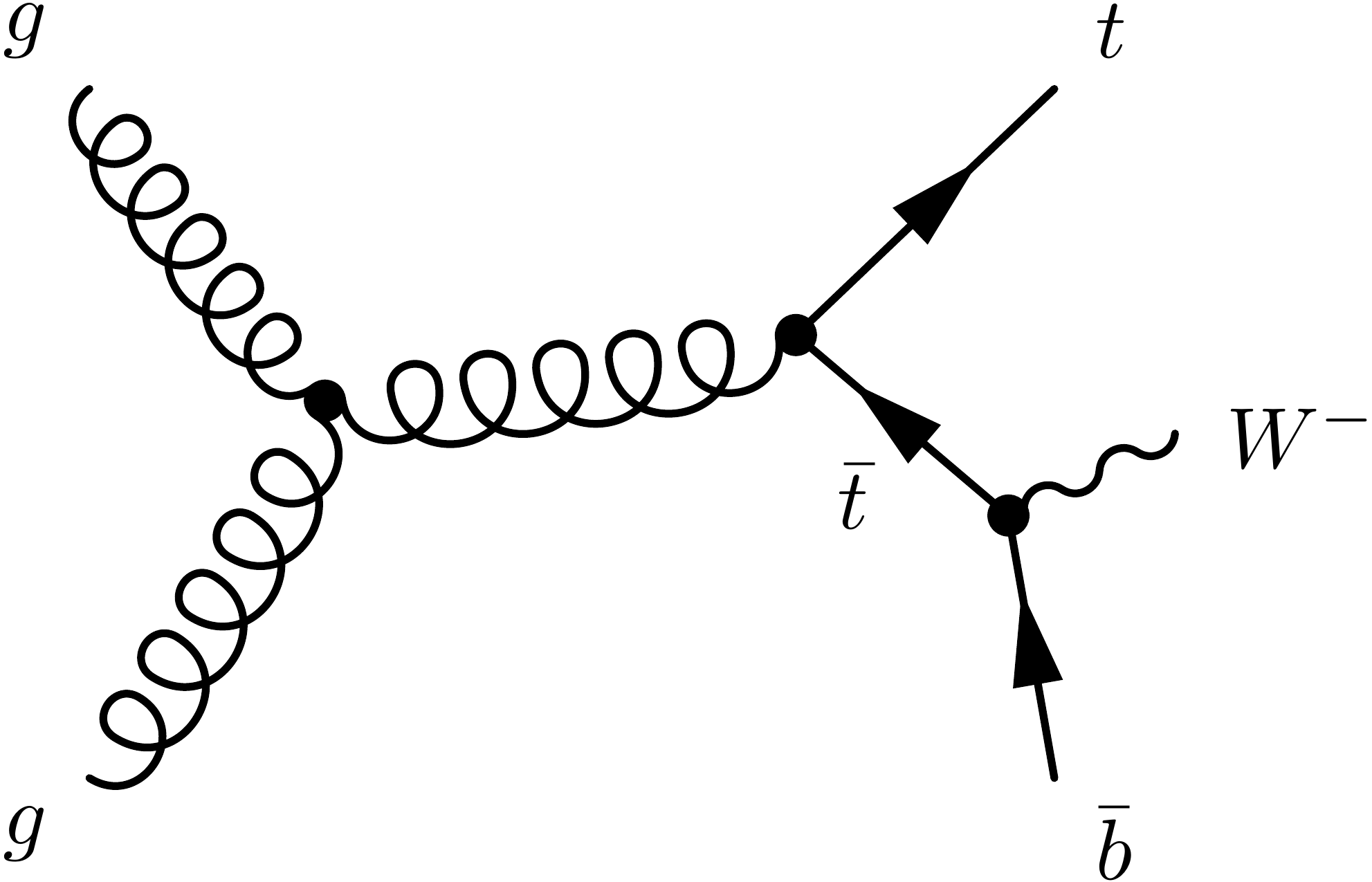}
	    \includegraphics[width=0.3\columnwidth]{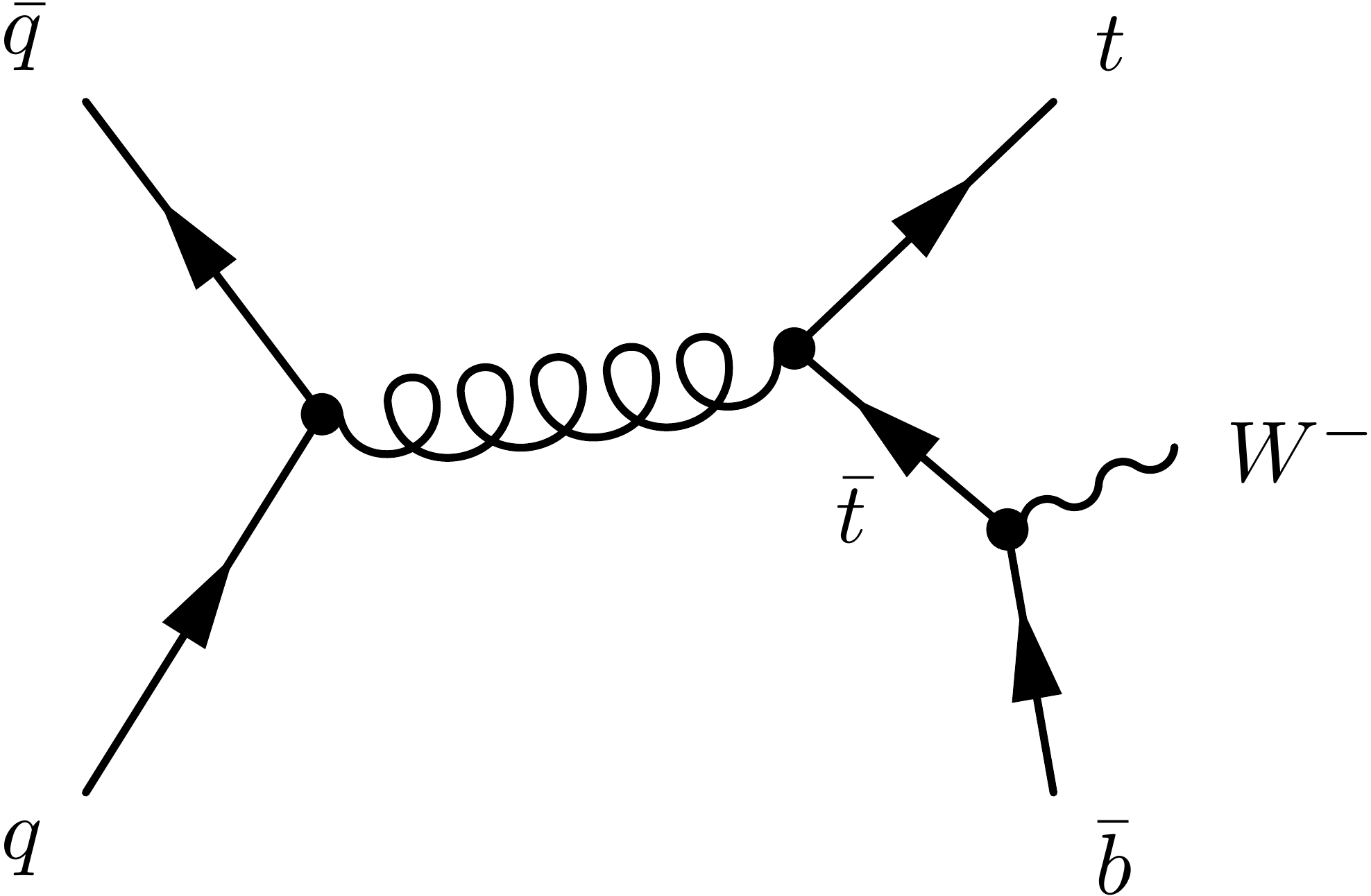}
	    \caption{\label{fig:fdnlo}Feynman diagrams for $\cPqt\PW$ single-top-quark production at next-to-leading order that are removed from the signal definition in the DR scheme, the charge-conjugate modes are implicitly included.}
	  \end{center}
	\end{figure}

The leading order Feynman diagrams for $\cPqt\PW$ production are shown in Fig.~\ref{fig:fdlo}. The definition of $\cPqt\PW$ production in perturbative QCD mixes with top quark pair production (\ttbar) at next-to-leading order (NLO)~\cite{Belyaev:1998dn,White:2009yt}. Two schemes are proposed to describe the $\cPqt\PW$ signal: ``diagram  removal" (DR)~\cite{Frixione:2008yi}, where all NLO diagrams which are doubly resonant, such as those in Fig.~\ref{fig:fdnlo}, are excluded from the signal definition; and ``diagram subtraction" (DS)~\cite{Frixione:2008yi,Tait:1999cf}, in which the differential cross section is modified with a gauge-invariant subtraction term, that locally cancels the contribution of \ttbar diagrams. The DR scheme is used in this Letter, but it has been verified that the number of predicted events after full selection is consistent between the two approaches within the statistical uncertainties of the simulated samples. The differences are accounted for in the systematic uncertainties.

In the standard model, top quarks decay almost exclusively to a $\PW$ boson and a $\cPqb$ quark. The study presented here has been performed in the channels in which both $\PW$ bosons decay leptonically into a muon or an electron and a neutrino, with a branching fraction ${\cal B}(\PW \to \ell\nu) = (10.80 \pm 0.09)$\%, where $\ell = \Pe$ or $\mu$~\cite{pdg}. The dilepton final states of the $\cPqt\PW$ process are characterized by the presence of two isolated leptons with opposite charge, a jet from the fragmentation of a $\cPqb$ quark, and a substantial amount of missing transverse energy (\MET) due to the presence of the neutrinos. The primary source of background events arise from \ttbar production, followed by \DY+jets processes.

The analysis uses fits to a discriminant variable built from kinematic quantities combined with a multivariate technique. A second analysis, intended as a cross-check of the robustness of the selection, is performed using event counting. In both cases, a sample collected at $\sqrt{s}$ = 7\TeV by CMS, corresponding to an integrated luminosity of 4.9\fbinv, is used.

The central feature of the CMS apparatus is a superconducting solenoid of 6\unit{m} internal diameter, providing a magnetic field of 3.8\unit{T}. Within the field volume are a silicon pixel and strip tracker, a lead tungstate crystal electromagnetic calorimeter, and a brass/scintillator hadron calorimeter. Muons are measured in gas-ionization detectors embedded in the steel return yoke. Extensive forward calorimetry complements the coverage provided by the barrel and endcap detectors. A more detailed description can be found in Ref.~\cite{CMS:2008zzk}. %The CMS experiment uses a right-handed coordinate system, with the origin at the nominal interaction point, the $x$ axis pointing to the center of the LHC ring, the $y$ axis pointing up (perpendicular to the to the plane of the LHC ring), and the $z$ axis along the counterclockwise-beam direction. The polar angle $\theta$ is measured from the $z$ axis and the azimuthal angle $\phi$ is measured in the $x$-$y$ plane. The pseudorapidity $\eta$ is defined as $\eta = -\ln [\tan(\theta/2)]$.

Single-top-quark events in all channels have been simulated with the \POWHEG event generator version 301~\cite{Frixione:2007vw}, designed to describe the full NLO properties of these processes, while \MADGRAPH 5.1.1~\cite{Alwall:2011uj} is used for \ttbar and for the inclusive single-boson production (V+X), where ${V=\PW,\cPZ}$ and X can indicate light or heavy partons. The remaining background samples are simulated using \PYTHIA version 6.4.24~\cite{pythia}, including diboson production and QCD multijet production enriched in events with electrons or muons produced in the decay of $\cPqb$ and $\cPqc$ quarks, and muons from the decay of long-lived hadrons. The CTEQ~6.6M pdf sets~\cite{cteq6} are used for all simulated samples. All generated events undergo a full simulation of the detector response using \GEANTfour~\cite{Agostinelli:2002hh,geant}. The value used for the top-quark mass is \mtop=172.5\GeV.

Approximate NNLO theoretical predictions are used to normalize \ttbar production \cmsBreak($\sigma_{\ttbar} = 163^{+11}_{-10}\unit{pb}$)~\cite{Kidonakis:2010dk}, \PW+jets and \DY+jets processes are normalized to complete NNLO calculations for the inclusive cross sections, and NLO cross sections are used for diboson processes~\cite{Gavin:2010az}. Unless otherwise stated, the theoretical values of the cross section have been used in this Letter to normalize the simulation in figures and tables.

Leptons, jets and \MET are reconstructed by the CMS particle flow (PF) algorithm~\cite{bib:particleFlow}, which performs a global event reconstruction and provides the full list of particles identified as electrons, muons, photons, and charged and neutral hadrons.

Events are collected using dilepton triggers with electrons or muons. The lepton transverse energy thresholds are symmetric, the highest used in these triggers is 17\GeV while the lowest is 8\GeV. The two selected leptons must originate from the same primary vertex and have opposite charge. The primary vertex used is defined as the reconstructed vertex with the highest \pt of associated tracks and is required to have at least four tracks, with longitudinal (radial) distance of less than 24 (2) cm from the center of the detector. Muon (electron) candidates are required to have a transverse momentum $\pt>20\GeV$ and pseudorapidity $|\eta|<2.4\,(2.5)$; events with additional leptons passing looser quality criteria are vetoed.

To remove low invariant mass Drell--Yan (\DY) events, the invariant mass of the lepton pair ($m_{\ell\ell}$) is required to be greater than 20\GeV. In the \Pe\Pe\ and $\mu\mu$ final states, events are also rejected if $m_{\ell\ell}$ is between 81 and 101\GeV, compatible with the \cPZ~boson mass; this veto removes background from \DY+jets, as well as from \cPZ\cPZ\ and \PW\cPZ\ processes. In the \Pe\Pe~and $\mu\mu$ decay channels, a requirement is applied on the \MET as well to further reduce the contribution from events without genuine \MET (mostly \DY+jets and QCD multijet production). Since the \MET resolution is degraded in events with high pileup, an additional quantity is used (tracker-\MET), calculated using only the charged particles associated with the primary vertex. Events are selected if both \MET and tracker-\MET are larger than 30\GeV.

Jets are defined according to the anti-$k_{\mathrm{T}}$ algorithm~\cite{Cacciari:2008gp} with a distance parameter of 0.5. Jets within $|\eta|<2.4$ and with \pt$>$30\GeV are considered in the analysis.

Exactly one jet is required to be present in the event, and it must be identified as coming from a $\cPqb$ quark. The identification of $\cPqb$ jets is done according to an algorithm that reconstructs the secondary vertex of the decay of the $\cPqb$ quark~\cite{btaggingPAS2010,btaggingPAS2011}, resulting in a discriminating variable sensitive to the lifetime of $\cPqb$ hadrons. The selection on this discriminant yields a \cPqb-tagging efficiency of 62\% with a mistag rate of 1.4\% for jets with \pt between 50 and 80\GeV.  Events with additional \cPqb-tagged jets with $\pt>20\GeV$ are removed. After this selection, the sample is dominated by \ttbar events and $\cPqt\PW$ signal.

Additionally, events with exactly two jets, in which either one or both jets have been $\cPqb$ tagged, are used in the fit. Three regions are defined per dilepton final state: one region with one jet that is $\cPqb$ tagged (1j1t) where the $\cPqt\PW$ signal is substantial, and two regions with two jets, where the \ttbar background is dominant, and exactly one or two $\cPqb$ tags are required (2j1t and 2j2t, respectively).

A smaller background comes from \DY events. It is found that in high-pileup scenarios the \MET distribution for \DY events is not properly modeled by the simulation, leading to disagreement between data and simulation. To solve this problem, the \DY simulation is corrected to match the missing transverse energy distribution observed in the data using events from the Z resonance.

The contributions of other backgrounds, i.e., diboson production (\PW\PW, \PW\cPZ, \cPZ\cPZ), QCD, \PW+jets, and other single-top-quark processes, are small, less than 1\% of the selected events, and estimated from simulation.

\begin{table}[h]
\topcaption{Event yields in the different regions. The simulation is quoted with statistical (first) and systematic uncertainties (second). When only one uncertainty is quoted, it is the total one.
\label{tab:selevents}}
\scotchrule[@{}lccc@{}]
 & 1j1t & 2j1t &  2j2t  \\
\hline
\cPqt\PW                                      &  336$\pm$5$\pm$16            	& 180$\pm$3$\pm$16             		& 45$\pm$1$\pm$6 \\
\ttbar                                 &  1263$\pm$19$\pm$138	         &  2775$\pm$28$\pm$205  		& 1488$\pm$21$\pm$222 \\
\DY+jets                           &   128$\pm$12$\pm$28          	& 113$\pm$10$\pm$22    			&  8.5$\pm$1.8$\pm$1.8   \\
Other                   	     &   19$\pm$3            			& 8.8$\pm$0.7$\pm$0.2           		& 4$\pm$3 \\
\hline
Total estimated        	      & 1746$\pm$23$\pm$141   	& 3077$\pm$30$\pm$207   		& 1546$\pm$21$\pm$222 \\
Total data                         & 1699                          			& 2878                				& 1507\\
\donescotchrule
\end{table}

The number of events in the signal and two control regions is presented for data and simulation in Table \ref{tab:selevents}. The approximate composition of the sample at this level is 70\% \ttbar events with 20\% $\cPqt\PW$ events in the signal region. In the 2j1t region the \ttbar content represents 90\% of the events, while $\cPqt\PW$ events are less than 6\%. In the 2j2t region, more than 95\% of the events are \ttbar events.

        \begin{figure}[h]
          \begin{center}
            \includegraphics[width=0.48\textwidth]{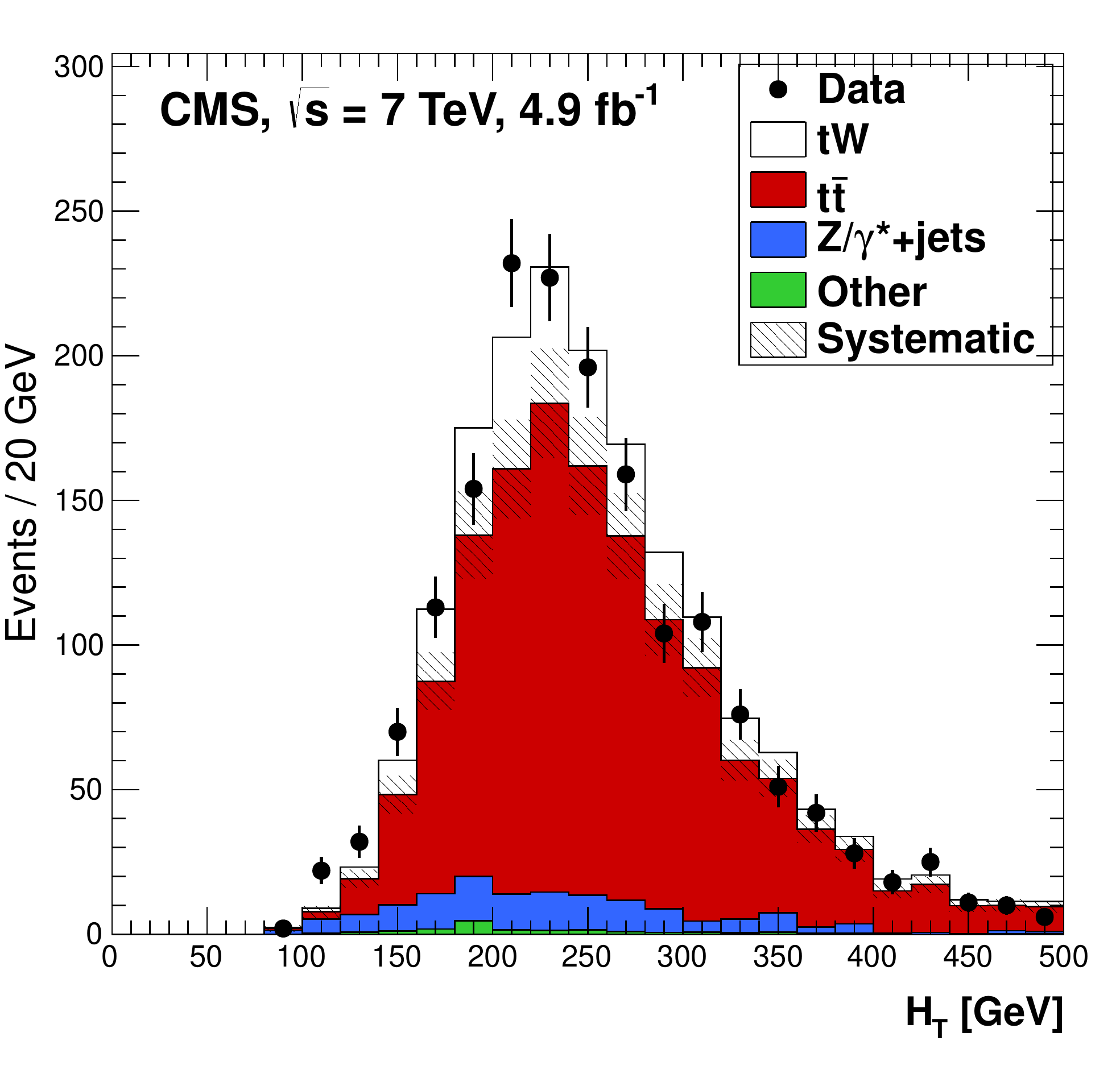}
            \includegraphics[width=0.48\textwidth]{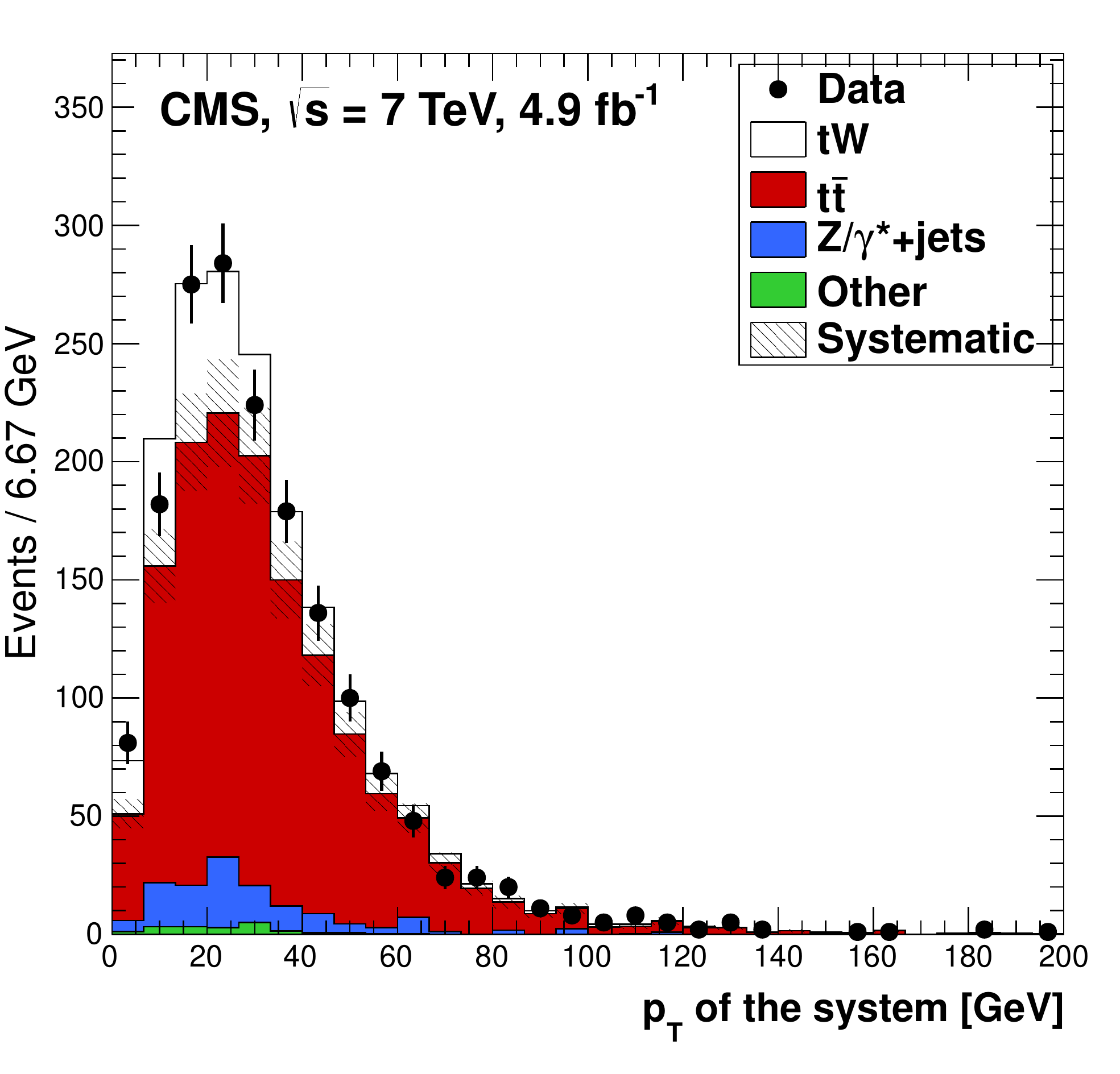}
            \caption{\label{fig:ptsysht}  Distributions  of $H_T$ and the \pt of the system composed of the leptons, \MET and the jet, in data and simulation after jet selection in the signal region (1j1t).}
          \end{center}
        \end{figure}

A multivariate analysis based on boosted decision trees (``BDT"  analysis)~\cite{decision_trees,Freund1997119} is used, testing the overall compatibility of the signal event candidates with the event topology of the $\cPqt\PW$ associated production.  Four variables are chosen to train the BDT based on their ability to separate the $\cPqt\PW$ signal from the dominant \ttbar background. These variables are $H_{\mathrm{T}}$, defined as the scalar sum of the transverse momenta of the leptons, jet, and \MET; the \pt of the system composed of the leptons, \MET and jet; the \pt of the jet with the highest energy; and the difference in angular separation, $\phi$, between the direction associated to the \MET and the closest of the two selected leptons. The distributions of $H_{{\mathrm T}}$ and the \pt of the system composed of the leptons, \MET and the jet, are presented, in the signal region (1j1t), in Fig.~\ref{fig:ptsysht}. The presence of the $\cPqt\PW$ signal over the background is visible in all the distributions. The distributions of the other two variables are available in \suppMaterial.

	\begin{figure}[h]
	  \begin{center}
	    \includegraphics[width=0.48\textwidth]{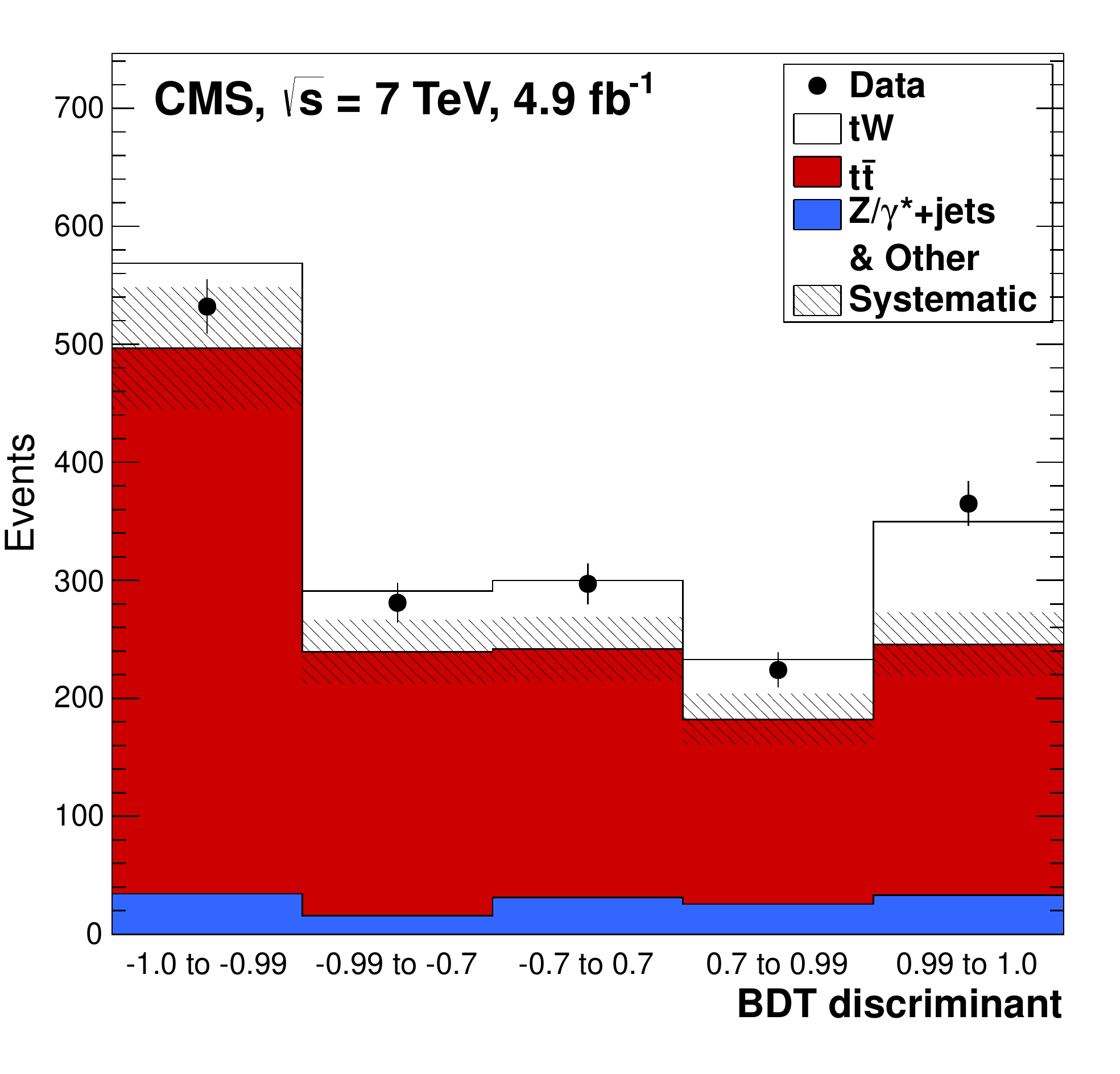}
		    \caption{\label{fig:bdtplots}  Distribution of the BDT discriminant in the signal region (1j1t) in data and simulation.}
	  \end{center}
	\end{figure}

The output of the BDT is a single discriminant value for every event ranging from $-1$ (background-like) to $+1$ (signal-like).  The distribution of the BDT discriminant is shown for the 1j1t signal region in Fig.~\ref{fig:bdtplots}. Even if
the $\cPqt\PW$ signal does not peak strongly at $+1$, its distribution discriminates it with respect to \ttbar and other backgrounds. Maximum signal sensitivity is achieved through a simultaneous fit to 9 categories: the three BDT discriminant shapes (1j1t, 2j1t, and 2j2t) in the three final states (\Pe\Pe,~\Pe$\mu$~and $\mu\mu$). The two \ttbar enriched regions are included to control the rate of this background in the signal region.

The impact of each individual source of uncertainty on the analysis has been estimated in every region and final state. The dominant systematic uncertainty that affects the rate of the $\cPqt\PW$ signal is associated with the \cPqb-tagging efficiency, with values between 3\% and 6\% for the different final states. The \cPqb-tagging efficiency uncertainty is also important for the \ttbar background yield, with values between 1.5\% and 4.0\%. The main systematic uncertainty for the \ttbar background is due to the factorization/renormalization scale used in the simulation, up to 11\%, with values around 2\% for $\cPqt\PW$ signal. Also for \ttbar, the uncertainties due to jet energy scale (7\%) and the threshold used to match the matrix element generator to the parton shower model in simulation (3\%) are important. The statistical uncertainty is the largest contribution to the uncertainty of the measured cross section, with a 20\% effect. The complete information about the systematic uncertainties is available in tabulated form in \suppMaterial.

A binned likelihood fit is performed on the distributions of the BDT discriminant. Template shapes for the signal and backgrounds are taken from simulation. Distributions are included separately in the fit for each of the three dilepton channels (\Pe\Pe, \Pe\Pgm, and \Pgm\Pgm) in the signal region (1j1t) and control regions (2j1t and 2j2t). Signal and background rates are allowed to vary in the fit, using the systematic uncertainties on the background rates as constraint terms in the likelihood function. The signal rate and 68\% confidence level (CL) interval is determined using the profile likelihood method. The sources of theoretical uncertainty that affect the template shape are then considered. For each uncertainty, ${\pm}1\,\sigma$ systematic shifts are applied to the simulated samples to obtain revised templates. Differences in signal rate found using the revised templates are taken as systematic uncertainties and are added in quadrature to the $1\,\sigma$ interval from the fit using the baseline templates. The expected significance is evaluated using the median and central 68\% of the values obtained from pseudo-experiments generated using the theoretical prediction of the standard model $\cPqt\PW$ cross section.
	
An excess of events over the expected background is observed with a significance of 4.0 $\sigma$, compatible with the expected significance of the $\cPqt\PW$ signal, $3.6^{+0.8}_{-0.9}\,\sigma$. The measured cross section, including both statistical and systematic uncertainties, is $16^{+5}_{-4}~\text{pb}$, in agreement with the standard model prediction. %The measurement is limited by the statistical uncertainty. Other substantial contributions to the uncertainties include uncertainties from the jet energy scale and signal modeling.

The measurement can be used to determine the absolute value of the Cabibbo-Kobayashi-Maskawa matrix element $|V_{{\cPqt\cPqb}}|$, following the same technique as in~\cite{Chatrchyan:1359133}, assuming that $|V_{{\cPqt\cPqd}}|$ and $|V_{{\cPqt\cPqs}}|$ are much smaller than $|V_{{\cPqt\cPqb}}|$:
\begin{equation} |V_{{\cPqt\cPqb}}| = \sqrt{\frac{\sigma_{\cPqt\PW}}{\sigma_{\cPqt\PW}^{{\text{th}}}}} = 1.01^{+0.16}_{-0.13}(\text{exp.}) ^{+0.03}_{-0.04}(\text{th.})\end{equation}
where $\sigma_{{\cPqt\PW}}^\text{th}$ is the standard model prediction computed assuming $|V_{\cPqt\cPqb}|=1$. Using the standard model assumption of $0\le |V_{\cPqt\cPqb}|^2 \le 1$, a value of $|V_{\cPqt\cPqb}|=1.00$ is inferred, with a 90\% confidence level interval of [0.79,1.00]. This is based on profile likelihood intervals, the same method used for the cross section measurement and intervals. Studies with pseudo-experiments were performed, showing the validity of the profile likelihood method in presence of the boundary $|V_{\cPqt\cPqb}|\leq1.0$.

	\begin{figure}[h]
	  \begin{center}
	    \includegraphics[width=0.48\textwidth]{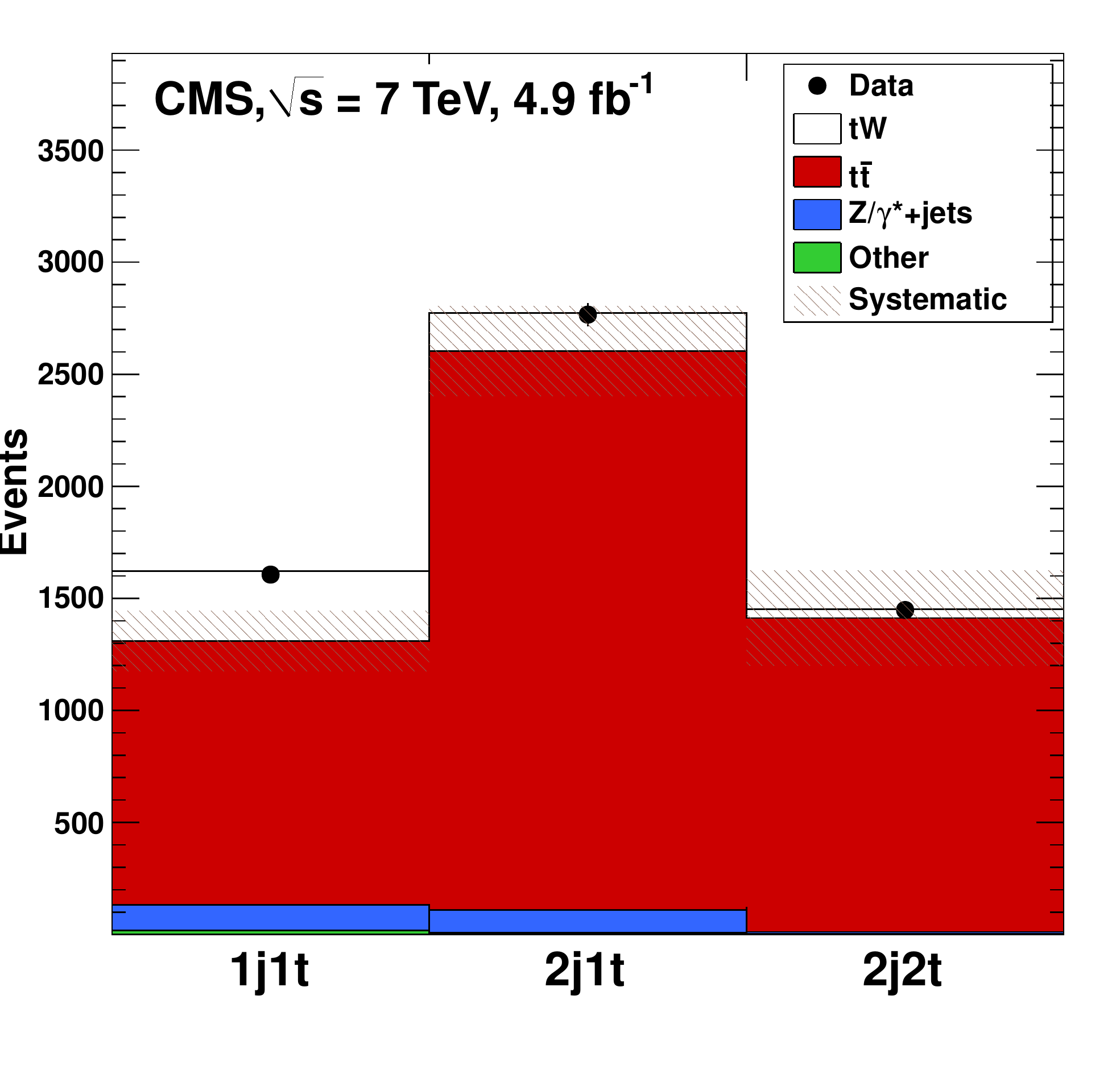}
               \caption{\label{fig:ttcr} Event yields in data and simulation in the signal region (1j1t) and the two \ttbar-enriched control regions for the count-based analysis. Simulation yields are scaled to the outcome of the fit.  }
	  \end{center}
	\end{figure}
	
A second analysis (``count-based"  analysis), used as a cross-check, is performed using event counts. After the jet selection step, instead of building the BDT discriminant, events are required in addition to have $H_{\mathrm{T}}>60\GeV$ in the \Pe\Pgm\ channel, where no invariant mass and \MET requirements are applied. The analysis uses a statistical model of Poisson event counts in the three dilepton final states in the signal region (1j1t) and control regions (2j1t and 2j2t).The event yield for each process in every region is affected by different sources of systematic uncertainties, equivalent to the ones calculated for the BDT analysis. These are included in the model as nuisance parameters. The same methods for the cross section measurement and the significance calculation as in the BDT analysis have been used. Figure~\ref{fig:ttcr} shows the event yields selected by the count-based analysis for each region, in data and simulation, in which the simulation yields have been normalized to the outcome of the maximum likelihood fit. The observed significance of the $\cPqt\PW$ signal obtained with the count-based analysis is $3.5\,\sigma$, with an expected significance of $3.2\pm0.9\,\sigma$. The count-based analysis measures a cross section of $15\pm5\unit{pb}$. These results are consistent with those obtained with the BDT analysis.

In summary, using 4.9\fbinv of data collected with the CMS experiment at the LHC, evidence has been found for the associated production of a single top quark and $\PW$ boson in \Pp\Pp\ collisions at $\sqrt{s}$ = 7\TeV with a significance of $4.0\,\sigma$ and a measured cross section of $16^{+5}_{-4}\unit{pb}$.

We congratulate our colleagues in the CERN accelerator departments for the excellent performance of the LHC machine. We thank the technical and administrative staff at CERN and other CMS institutes, and acknowledge support from: BMWF and FWF (Austria); FNRS and FWO (Belgium); CNPq, CAPES, FAPERJ, and FAPESP (Brazil); MES (Bulgaria); CERN; CAS, MoST, and NSFC (China); COLCIENCIAS (Colombia); MSES (Croatia); RPF (Cyprus); MoER, SF0690030s09 and ERDF (Estonia); Academy of Finland, MEC, and HIP (Finland); CEA and CNRS/IN2P3 (France); BMBF, DFG, and HGF (Germany); GSRT (Greece); OTKA and NKTH (Hungary); DAE and DST (India); IPM (Iran); SFI (Ireland); INFN (Italy); NRF and WCU (Korea); LAS (Lithuania); CINVESTAV, CONACYT, SEP, and UASLP-FAI (Mexico); MSI (New Zealand); PAEC (Pakistan); MSHE and NSC (Poland); FCT (Portugal); JINR (Armenia, Belarus, Georgia, Ukraine, Uzbekistan); MON, RosAtom, RAS and RFBR (Russia); MSTD (Serbia); SEIDI and CPAN (Spain); Swiss Funding Agencies (Switzerland); NSC (Taipei); TUBITAK and TAEK (Turkey); STFC (United Kingdom); DOE and NSF (USA).

\bibliography{auto_generated}   % will be created by the tdr script.
\ifthenelse{\boolean{cms@external}}{}{
\clearpage
\appendix
\input{additional}

}

\cleardoublepage \appendix\section{The CMS Collaboration \label{app:collab}}\begin{sloppypar}\hyphenpenalty=5000\widowpenalty=500\clubpenalty=5000\input{TOP-11-022-authorlist.tex}\end{sloppypar}
\end{document}

%% file: additional.tex
\section{Supplemental Information}\label{app:suppMat}

\begin{figure}[htb]
          \begin{center}
            \includegraphics[width=0.48\textwidth]{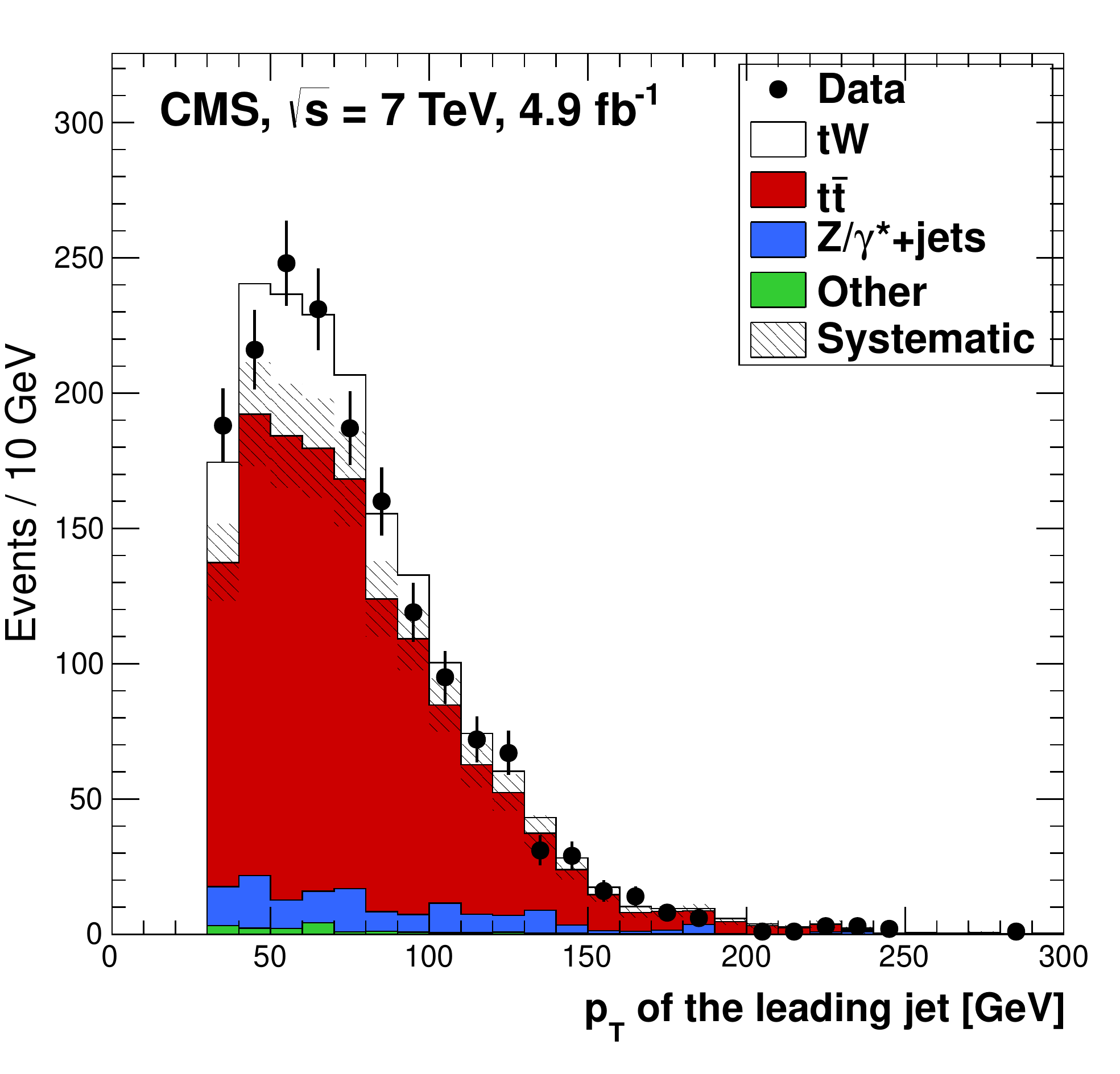}
            \includegraphics[width=0.48\textwidth]{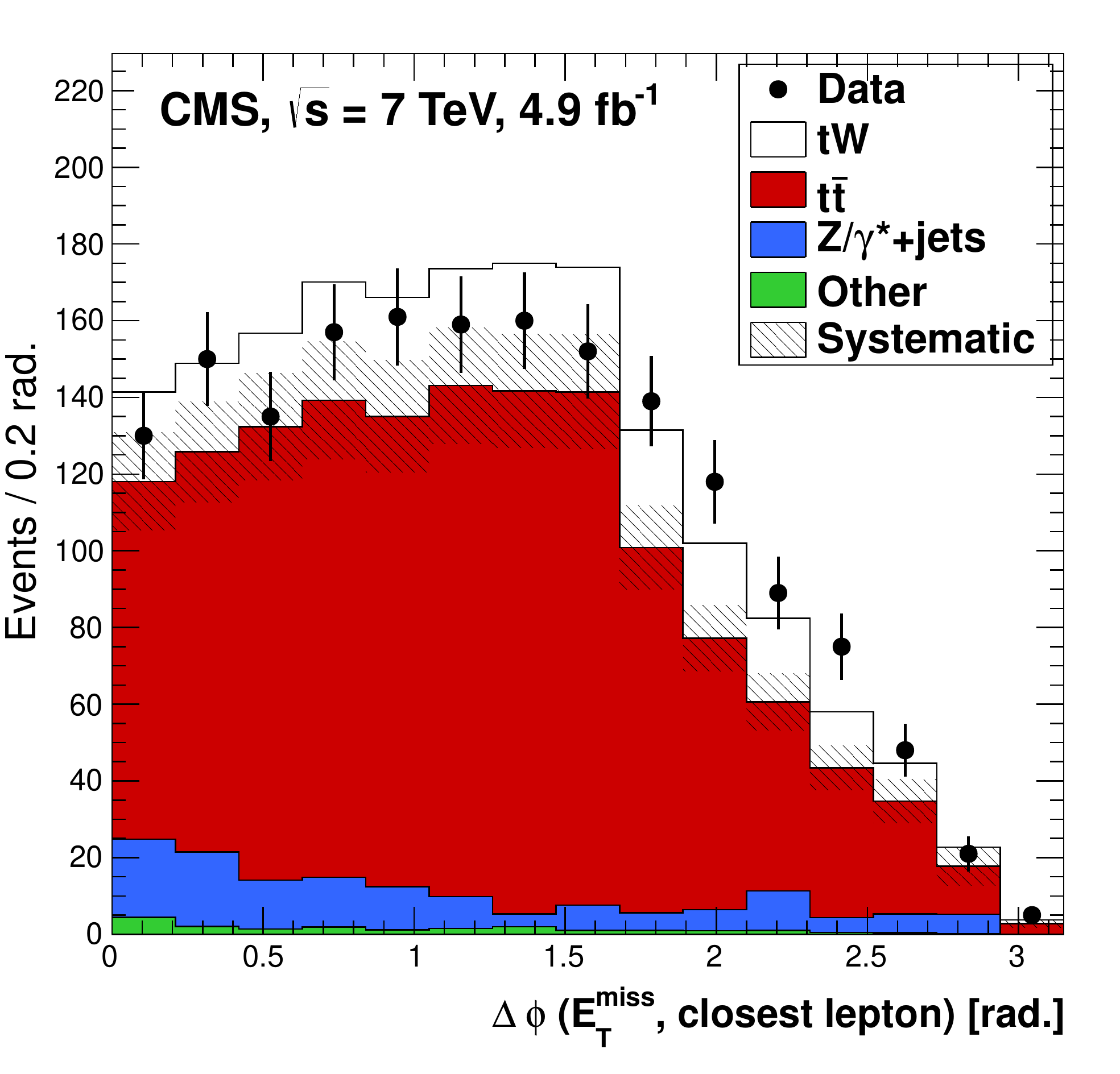}
            \caption{\label{fig:ptphi}  Distributions  of the $\pt$ of the jet with the highest energy and the difference in $\phi$ between the direction associated to the $\mathrm{E}_{\mathrm{T}}^{\mathrm{miss}}$ and the closest lepton in data and simulation after jet selection in the signal region (1j1t).}
          \end{center}
       \end{figure}

This document presents additional material to the publication. Figure \ref{fig:ptphi} shows the distributions of the two variables of the BDT not presented before, and Tables \ref{tab:add1} and \ref{tab:add2} contain information related to the systematic uncertainties that affect the analysis.

The distributions of the $p_{\mathrm{T}}$ of the jet with the highest energy and the difference in angular separation, $\phi$, between the direction associated to the $\mathrm{E}_{\mathrm{T}}^{\mathrm{miss}}$ and the closest of the two selected leptons, in data and simulation, are presented in the signal region (1j1t) in Fig.~\ref{fig:ptphi}.

Table \ref{tab:add1} presents the impact of each individual source of systematic uncertainty on the rate of the different processes in the signal region for each final state. When two numbers are listed for a single uncertainty, the upper number is the effect on the rate when the systematic uncertainty source is scaled up and the lower is for when it is scaled down. Entries marked with a ``-" either do not apply for that particular final state/process, or have a negligible effect. Other processes refers to \cPZ/$\gamma^\star$ and the rest of backgrounds, that have almost negligible contributions.

\begin{table*}[htb]
\topcaption{Impact of each individual source of systematic uncertainty on the rate of the different processes in the signal region for each final state. Other processes refers to \cPZ/$\gamma^\star$ and the rest of backgrounds, that have almost negligible contributions.
\label{tab:add1}}
\scotchrule[ l  c  c  c ]
Systematic Uncertainty ($\Pe\Pe/\Pe\Pgm/\Pgm\Pgm$) & \cPqt\PW & $\ttbar$ & Other processes\\
& (\%) & (\%) & (\%) \\
\hline
Luminosity & ${2.2}$/${2.2}$/${2.2}$ & ${2.2}$/${2.2}$/${2.2}$ & ${2.2}$/${2.2}$/${2.2}$ \\
Pileup modeling& ${2.1}$/${0.6}$/${0.2}$ & ${1.0}$/${0.5}$/${0.3}$ & ${4.9}$/${1.6}$/${3.7}$ \\
Electron Trigger Efficiency & ${1.5}$/${1.1}$/- & ${1.5}$/${1.1}$/- & ${1.5}$/${1.1}$/- \\
Muon Trigger Efficiency & -/${1.1}$/${1.5}$ & -/${1.1}$/${1.5}$ & -/${1.1}$/${1.5}$ \\
Electron Identification & ${2}$/${2}$/- & ${2}$/${2}$/- & ${2}$/${2}$/- \\
Muon Identification & -/${1}$/${1}$ & -/${1}$/${1}$ & -/${1}$/${1}$ \\
b-tagging &  $^{+2.6}_{-4.0}$/$^{+4.1}_{-3.7}$/$^{+4.1}_{-3.3}$ & $^{+3.2}_{-1.1}$/$^{+2.5}_{-3.3}$/$^{+2.6}_{-3.4}$ & $^{+3.9}_{-7.0}$/$^{+3.4}_{-3.6}$/$^{+3.2}_{-3.5}$ \\
Jet Energy Scale & $^{-1.1}_{+0.4}$/$^{-0.6}_{+1.7}$/$^{-1.8}_{+2.1}$ & $^{-4.3}_{+7.6}$/$^{-4.7}_{+3.7}$/$^{-3.6}_{+5.5}$ & $^{+13.2}_{+1.7}$/$^{-1.2}_{-0.1}$/$^{+11.2}_{+10.7}$ \\
Jet Energy Resolution & $^{+1.2}_{+1.4}$/$^{+0.1}_{+0.2}$/$^{+0.2}_{+0.8}$ & $^{+2.9}_{+0.5}$/$^{-0.5}_{-2.2}$/$^{+2.9}_{+0.1}$ & $^{+8.0}_{-5.3}$/$^{-4.4}_{+7.7}$/$^{+1.4}_{-11.9}$ \\
$\mathrm{E}_{\mathrm{T}}^{\mathrm{miss}}$ modeling & $^{-0.6}_{+1.0}$/-/$^{+1.0}_{-0.2}$ & $^{+0.2}_{-0.9}$/-/$^{+0.3}_{+0.5}$ & $^{+2.4}_{-7.7}$/$^{-0.1}_{+0.2}$/$^{+13.5}_{-2.5}$ \\
Factorization and Normalization scale $\mathrm{Q^2}$ & $^{+3.1}_{+0.4}$/$^{+3.4}_{+0.3}$/$^{+3.3}_{+0.4}$ & $^{+10.0}_{-6.5}$/$^{+10.1}_{-6.9}$/$^{+10.1}_{-6.9}$ & -/-/- \\
ME/PS Matching Thresholds &  -/-/- & $^{+0.7}_{+2.0}$/$^{+0.5}_{+1.6}$/$^{+0.4}_{+1.8}$ & -/-/- \\
tW DR/DS scheme & $^{-5.4}_{-0.5}$/$^{+6.4}_{-0.5}$/$^{+1.7}_{-0.5}$ & -/-/- & -/-/- \\
PDF uncertainties & ${2.2}$/${2.0}$/${2.0}$ & -/-/- & -/-/- \\
 $\ttbar$ cross-section &  -/-/- & $^{+6.2}_{-5.8}$/$^{+6.2}_{-5.8}$/$^{+6.2}_{-5.8}$ & -/-/- \\
\cPZ/$\gamma^\star$ modeling & -/-/- & -/-/- & ${30.5}$/${1.2}$/${23.5}$ \\
Simulation Statistics & ${3.8}$/${1.8}$/${2.7}$ & ${4.5}$/${2.0}$/${2.9}$ & ${18.0}$/${12.0}$/${12.4}$ \\
\hline
\donescotchrule
\end{table*}

Table \ref{tab:add2} presents the contribution to the uncertainty of the measured cross section of the different sources of uncertainty considered in the analysis. This is estimated by fixing each source one at a time and measuring the effect in the cross section uncertainty.

\begin{table}[htb]
\topcaption{Contribution to the uncertainty of the measured cross section of the different sources of uncertainty considered in the BDT analysis.
\label{tab:add2}}
\scotchrule[ l  c  c ]
  Systematic Uncertainty & $\Delta \sigma$ (pb) & $\frac{\Delta \sigma}{\sigma}$ \\
  \hline
  Luminosity                     &  0.69  & 0.04\\
  Pileup modeling                &  0.24  & 0.02\\
  Electron trigger efficiency    &  0.35  & 0.02\\
  Muon trigger efficiency        &  0.38  & 0.02\\
  Electron identification        &  0.70  & 0.04\\
  Muon identification            &  0.45  & 0.03\\
  b-tagging                      &  0.30  & 0.02\\
  Jet Energy Scale               &  2.42  & 0.15\\
  Jet Energy Resolution          &  0.58  & 0.04\\
$\mathrm{E}_{\mathrm{T}}^{\mathrm{miss}}$ modeling                  &  0.40  & 0.05\\
  tW $\mathrm{Q^2}$              &  0.34  & 0.02\\
    $\ttbar$ $\mathrm{Q^2}$         &  0.29  & 0.02\\
  ME/PS Matching Thresholds      &  1.62  & 0.10\\
  tW DR/DS scheme                &  0.94  & 0.06\\
  PDF  uncertainties             &  0.34  & 0.02\\
   $\ttbar$ cross section           &  0.96  & 0.06\\
  \cPZ/$\gamma^\star$ modeling                   &  0.67  & 0.04\\
  Statistical                    &  3.33  & 0.21\\
  \hline
  Total                          &  4.95  & 0.31\\
\donescotchrule
\end{table}

%% file: TOP-11-022-authorlist.tex
\textbf{Yerevan Physics Institute,  Yerevan,  Armenia}\\*[0pt]
S.~Chatrchyan, V.~Khachatryan, A.M.~Sirunyan, A.~Tumasyan
\vskip\cmsinstskip
\textbf{Institut f\"{u}r Hochenergiephysik der OeAW,  Wien,  Austria}\\*[0pt]
W.~Adam, E.~Aguilo, T.~Bergauer, M.~Dragicevic, J.~Er\"{o}, C.~Fabjan\cmsAuthorMark{1}, M.~Friedl, R.~Fr\"{u}hwirth\cmsAuthorMark{1}, V.M.~Ghete, J.~Hammer, N.~H\"{o}rmann, J.~Hrubec, M.~Jeitler\cmsAuthorMark{1}, W.~Kiesenhofer, V.~Kn\"{u}nz, M.~Krammer\cmsAuthorMark{1}, I.~Kr\"{a}tschmer, D.~Liko, I.~Mikulec, M.~Pernicka$^{\textrm{\dag}}$, B.~Rahbaran, C.~Rohringer, H.~Rohringer, R.~Sch\"{o}fbeck, J.~Strauss, A.~Taurok, W.~Waltenberger, G.~Walzel, E.~Widl, C.-E.~Wulz\cmsAuthorMark{1}
% ...
\vskip\cmsinstskip
\dag:~Deceased\\
1:~~Also at Vienna University of Technology, Vienna, Austria\\
% ...
